# Three steps towards dose optimization for oncology dose finding

Jason J.Z. Liao[a*], Ekaterine Asatiani[b], Qingyang Liu[a], Kevin Hou[a]

[a]Incyte Corporation, 1801 Augustine Cut-off, Wilmington, Delaware, 19803, USA;

jliao@incyte.com

[b]Incyte Biosciences International Sàrl, Rue Docteur-Yersin 12, 1110 Morges, Switzerland

**Abstract**

Traditional dose selection for oncology registration trials typically employs a one- or two-step single maximum tolerated dose (MTD) approach. However, this approach may not be appropriate for molecularly targeted therapy that tends to have toxicity profiles that are markedly different to cytotoxic agents. The US Food and Drug Administration launched Project Optimus to reform dose optimization in oncology drug development and has recently released a related Guidance for Industry. In response to these initiatives, we propose a "three steps towards dose optimization" procedure and discuss the details in dose optimization designs and analyses in this manuscript. The first step is dose-escalation to identify the MTD or maximum administered dose with an efficient hybrid design, which can offer good overdose control and increases the likelihood of the recommended MTD being close to the true MTD. The second step is the selection of appropriate recommended doses for expansion (RDEs), based on all available data including emerging safety, pharmacokinetics, pharmacodynamics, and other biomarker information. The third step is dose optimization, which uses data from a randomized fractional factorial design with multiple RDEs explored in multiple tumor cohorts during the expansion phase to ensure a feasible dose is selected for registration trials, and that the tumor type most sensitive to the investigative treatment is identified. We believe using this three-step approach can increase the likelihood of selecting the optimal dose for registration trial, one that demonstrates a balanced safety profile while retaining much of the efficacy observed at the MTD.

## 1. Introduction

Over the past decade, drug development in oncology has evolved and shifted from the use of cytotoxic agents to drugs with novel mechanisms of action (MOA), such as immunotherapies, targeted therapeutics, T-cell engagers, and others. Key differences exist in the MOAs, treatment procedures, pharmacodynamic and clinical effects among these therapies, which can markedly influence the dose optimization process for registration trials. As cytotoxic treatments are typically administered for short duration in fixed number of cycles and have narrow therapeutic indexes with steep dose-response/toxicity relationships, the maximum tolerated dose (MTD) is typically a reasonable dose for the registration study. In addition, serious toxicities from cytotoxic therapies are relatively predictable and often occur early in the treatment course. In contrast, the dose selection process for targeted therapeutics or immunotherapies can be much more complex; if tolerable, these treatments are usually administered until disease progression, which can be many months or years after study initiation. In the case of these noncytotoxic or selectively cytotoxic therapies, serious safety signals may only become apparent at later stages of treatment, and long-term toxicities above grade 2 may not be tolerated due to the chronic nature of these therapies. Moreover, targeted therapies may have wide therapeutic indexes and non-linear dose-response/toxicity relationships. Therefore, the MTD may not be reached during the course of the study and the optimal dose selected for the registration study may differ substantially from the MTD.

The traditional approach to defining a dose for a registration study typically employs a one- or two-step approach. (Shah, et al. 2021; 2022) The one-step approach involves conducting a dose-escalation study to determine the MTD of an investigative treatment, which is used as the study dose for comparison with standard of care (SOC) in subsequent registration trials. The two-step approach uses a dose-escalation study to determine the MTD, which is used in the expansion phase for different cohorts. Similar to the one-step approach, the MTD of the investigative treatment is then used as the registration study dose compared with the SOC. In the traditional one- or two-step approach, suboptimal characterization of the dosing schedule can lead to inappropriate dose selection for the registration trial, potentially leading to increased toxicity without additional efficacy. Severe toxicities may lead to high rates of dose reduction or premature discontinuation, resulting in failure to realize the full benefit of the investigative treatment. Furthermore, persistent or irreversible toxicities could potentially limit options for subsequent therapies and any benefits they may provide.

The old approach of "more is better" may be applicable for dose selection of chemotherapy treatment; however, this assumption no longer holds true for many of the newer, targeted therapies with vastly different MOAs and other features. (Shah, et al. 2021; 2022) Thus, the current paradigm for dose selection using the one- or two-step approach—developed around cytotoxic chemotherapeutics—may lead to investigative molecularly targeted therapies entering registration trials without adequately characterized dosing schedules. In response, the Oncology Center of Excellence (OCE) announced an initiative, Project Optimus, to reform the dose optimization and dose selection paradigm in oncology drug development. (FDA Oncology Center of Excellence, 2021; US Food and Drug Administration, 2022; Friends of Cancer Research, 2021) As a result, health authorities, in particular the US Food and Drug Administration (FDA), now mandate rigorous dose-finding and dose optimization processes before the initiation of pivotal trials for new oncology drugs. (Zirkelbach et al. 2022)

The goal of Project Optimus is to move forward with a dose-finding and dose optimization standard across oncology that emphasizes selection of a dose or doses to maximize not only the efficacy of a drug, but also safety and tolerability. The Project Optimus initiative recommends a balanced benefit/risk ratio in defining the optimal dose early in development. (Birri, et al., 2022; US Food and Drug Administration, 2022) Recently, the FDA published draft guidance for industry on optimizing dosage in the treatment of oncologic diseases, (FDA guidance, 2023) ensuring maximal efficacy is retained at optimal dose(s) relative to the MTD, while striving for a better balanced safety profile. This represents a shift towards identifying an optimal (biological) dose, which takes into account overall efficacy and tolerability where the MTD represents the upper limit of the optimal dose range, and away from solely determining the MTD. **Figure 1** illustrates the different relationships between the optimal dose and MTD/Maximum administered dose (MAD) among different drug treatments. Such optimization requires consideration of complex MOAs, schedule optimization, long-term drug tolerability, and potentially novel pharmacodynamic endpoints. Consequently, thoughtful study designs, exposure information, translational data, and statistical modeling play an increasingly important role. In response to Project Optimus and the FDA draft guidance on dose optimization in oncology drug development, we propose a "three-step towards dose optimization" procedure.

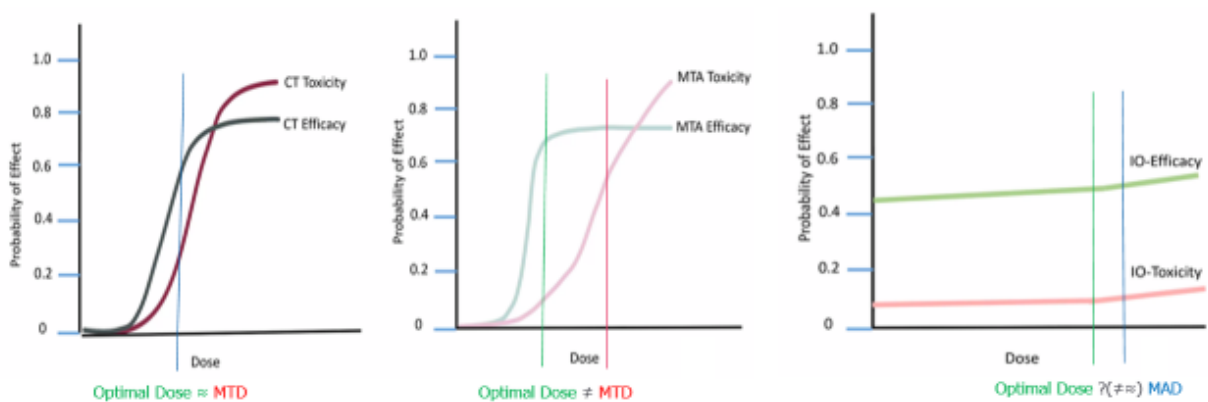

*Figure 1 Relationship between the optimal dose and MTD/MAD among different drug treatments. CT, cytotoxic therapy; IO, immunotherapy; MAD, maximum administered dose; MTA, molecularly targeted agent; MTD, maximum tolerated dose.*

## 2. Three steps towards dose optimization

Determining the optimal dose of a drug should start with considering the fundamentals of the therapeutic index. If the therapeutic window is wide, it is easier to develop a drug with a dose that satisfies a balanced efficacy and toxicity profile than if the therapeutic window is narrow. Regardless of the therapeutic index, it is important to first identify an upper boundary when searching for the optimal dose (**Figure 1**) as this narrows the search range for an optimal dose.

### 2.1: *Step 1: Dose-escalation in identifying an MTD/MAD*

Since the introduction of Project Optimus, there appears to be a misconception that the MTD of a non-cytotoxic drug is no longer relevant. However, finding an accurate MTD estimate that closely resembles the true MTD with a low likelihood of overdose toxicity remains pertinent. As described, the first step of searching for an optimal dose involves establishing the upper boundary to limit the search range. If the MTD of some targeted therapeutics or immunotherapies cannot be established, the MAD may be used as an alternative for establishing an upper boundary for an optimal dose.

The MTD is often identified during the dose-escalation part of the phase I study, with various study designs. We recommend a recently developed hybrid design (Liao et al. 2022) for identifying the MTD. This is a hybrid design on two levels, it is a hybrid of the modified mTPI design and a dose-toxicity model, and also a hybrid of the Bayesian approach for each individual dose level and the frequentist approach for combining available information from all tested doses. The design retains the merits of existing designs while minimizing the limitations. This hybrid design has demonstrated robust performance with good overdose control, which can minimize the difference in the recommended dose for trial and the true MTD. With the integration of all available dose groups in addition to a modified mTPI, the hybrid design could improve accuracy and efficiency in dose selection. The hybrid design is composed of three stages (**Figure 2**):

*Stage 1*: The mTPI design is modified to control the overdose toxicity more efficiently by an additional constraint using the posterior probability of the dose-limiting toxicity (DLT) rate in the overdosing interval ($\delta_2$, 1) being less than a value γ (e.g., <0.75).

*Stage 2*: This stage uses a dose-toxicity model by pooling observed safety information from all previous doses to estimate the DLT rate for the current dose level, and to predict the DLT rate for the next dose level in the provisional dose list.

*Stage 3*: Dose-escalation decisions from stages 1 and 2 are pooled to make a conservative dose-escalation decision to further control overdose toxicity.

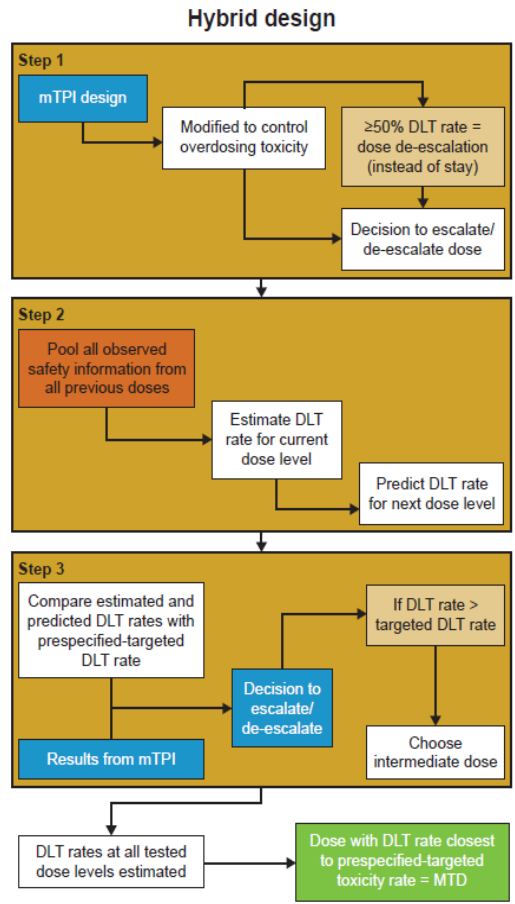

*Figure 2 Hybrid design for dose-escalation (Liao, et al. 2022). DLT, dose-limiting toxicity; MTD, maximum tolerated dose; mTPI, modified toxicity probability interval.*

Details of this hybrid design and an example of its application have been reported (Liao et al. 2022). To use this in practice, an R-package "*HybridDesign*" (Zhou, Zhou, and Liao, 2022) and an R-shiny tool (https://fzh223.shinyapps.io/HybridModel/) have been developed. These tools are freely available to guide clinicians at every step of the dose-finding process using the hybrid design.

As emphasized in Liao et al. (Liao et al. 2022), a dose-escalation procedure can be continued to search for an MTD with a tolerated toxicity limit. If the MTD is not reached or does not exist, but the MAD has satisfactory efficacy, the dose-escalation process may be stopped. In such cases, an upper boundary may be established using either MTD or MAD for efficient search of the optimal dose. To better characterize the exposure response before moving to the next step of identifying dose level for further evaluations, the backfill cohorts at certain dose levels in the dose-escalation study could be explored with more patients in selected populations to gain additional information on pharmacokinetics (PK) and pharmacodynamics (PD).

**2.2:** *Step 2: Dose selection to identify recommended doses for expansion (RDEs)*

After the MTD or MAD with satisfactory efficacy has been identified, multiple different dose levels may be selected for the dose-response assessment for efficacy and toxicity. The optimal dose can be difficult to identify, due to potential uncertainty of the response curve and variability at different dose levels. Therefore, multiple doses should be selected and recommended for expansion (phase II). (FDA guidance, 2023)

To identify the RDEs, all available data should be evaluated, this includes nonclinical and clinical data with emerging safety, PK, PD, and other biomarker information. This provides a preliminary understanding of dose- and exposure-response relationships for activity, safety, and tolerability. (FDA guidance, 2023) Prior to determining the RDEs, the dose/exposure and efficacy/toxicity-response relationships should be well-characterized. As the dose-exposure-response relationship shows in **Figure 3**, there are two calibration stages for deriving the dose for specified responses. The first stage is to derive the desired exposure from the specified responses (**Figure 4a**), where many modeling techniques can be applied. Some of the example model techniques, such as exposure model, PK/PD model, PK/safety model, PK/efficacy model, or biomarker/efficacy correlation, could be used to identify the RDEs. After the exposure information is explored, the second stage of determining RDEs is to implement the second calibration using exposure information from the first calibration to define the RDEs (**Figure 4b**). With MTD/MAD established, the RDEs should be no greater than MTD/MAD and no less than the pharmacologically active dose. As indicated in **Figure 1** with the relationship between the optimal dose and the MTD/MAD, the selections of RDEs for comparison with the MTD/MAD should build upon dose-response (efficacy/toxicity) and variability. Thus, it is desirable to have multiple RDEs for dose optimization with maximum benefit of a drug, which ideally includes not only longer survival but also an improved quality of life.

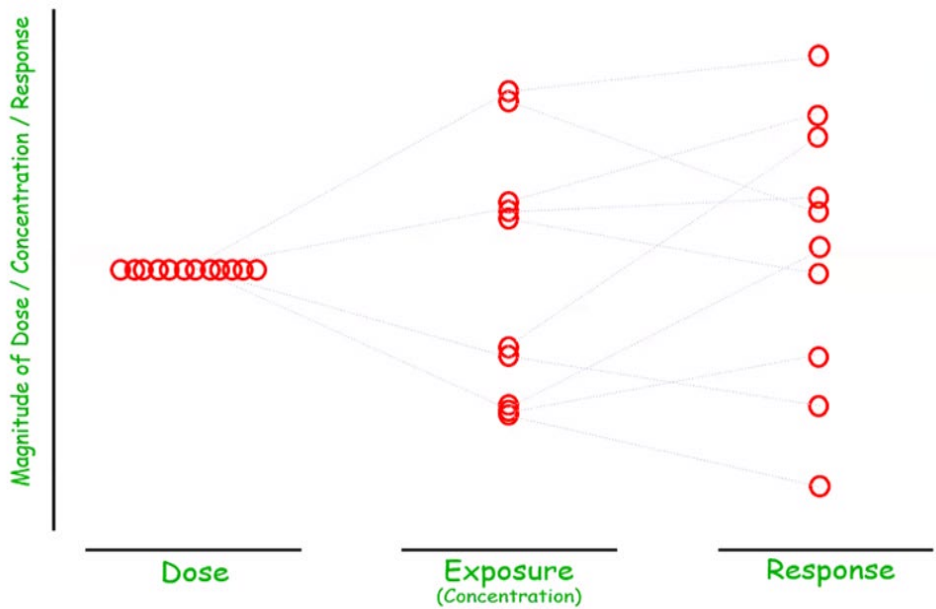

*Figure 3 The dose-exposure-response relationship.*

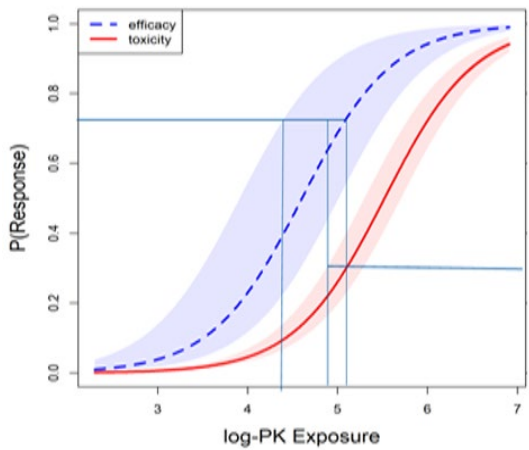 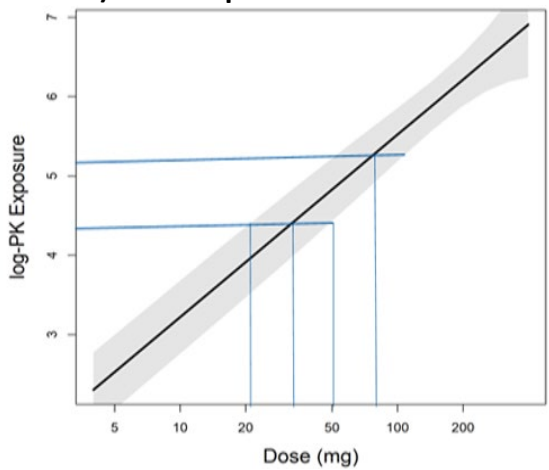

*Figure 4 The two calibration stages in deriving the RDEs: **(a)** identify the exposure range where the efficacy could be a selected biomarker, and **(b)** identify the doses. CI, confidence interval; PK, pharmacokinetics; P(Response): probability of response.*

As suggested in the 2023 FDA Guidance for Industry on dose optimization, multiple doses should be evaluated following dose-escalation for determining the optimal dose(s). The goal for evaluating multiple RDEs is to decrease uncertainty with identifying an optimal dosage(s); (FDA guidance, 2023) the chances of selecting the correct optimal dose(s) increase with the number of doses being evaluated. Araujo and colleagues suggested that at least two, but preferably three, RDEs to be selected for evaluation: one for the minimum reproducibly active dose, another for an effective dose close to the MTD, and one for an intermediate dose. (Araujo, et al. 2023) This suggestion is supported by the fact that at least three doses are required to accurately ascertain nonlinearity of the dose-response curve. Moreover, the greater the degree of nonlinearity, the more RDEs are needed to better capture the nonlinearity. In respect of selecting the optimal dose(s), three or more RDEs should be identified and a subset of these identified RDEs compared because of the potential nonlinearity of the dose-response and variability. The choice and number of RDEs selected for comparison should be balanced with scientific rigidity and with clinical and practical considerations.

**2.3: *Step 3: Dose optimization***

With selected RDEs in step 2 where the multiple doses are close to the optimal dose, a dose optimization procedure is conducted in a randomized, parallel, dose-response trial by applying the selected RDEs. (FDA guidance, 2023) At this early stage, more information is needed on tumor sensitivity, including that in different tumor types. In 2022, the US FDA issued guidance recommending the use of multiple expansion cohort trials to expedite oncology drug development. (FDA guidance, 2022) A unique feature of multiple expansion cohort trials is the two-dimensional dependency structure across doses and indications, which can assist with the identification of compound-sensitive tumors. Thus, the efficacy and safety of different RDEs are evaluated in different cohorts and an optimal dose is selected based on a balanced benefit-to-risk ratio through a randomized design. The design should be fit-for-purpose and could be a randomized multi-dose and/or multi-stage design, with options to discontinue inadequate dose arms to limit patient exposure to suboptimal or more toxic doses. (FDA guidance, 2023)

One such design involves randomizing patients in each cohort to all the selected RDEs. This commonly used full factorial design requires large numbers of patients with multiple different selected RDEs, which may impose a significant resource burden. To achieve dose optimization using a smaller number of patients and within shorter timelines, we propose a randomized fractional factorial design as illustrated in **Figure 5**. Within randomization in each cohort, two RDE doses are explored and a common dose used in each cohort; In addition to further characterizing exposure response relationship, this information can be used to identify the tumors most sensitive to treatment for a registration trial.

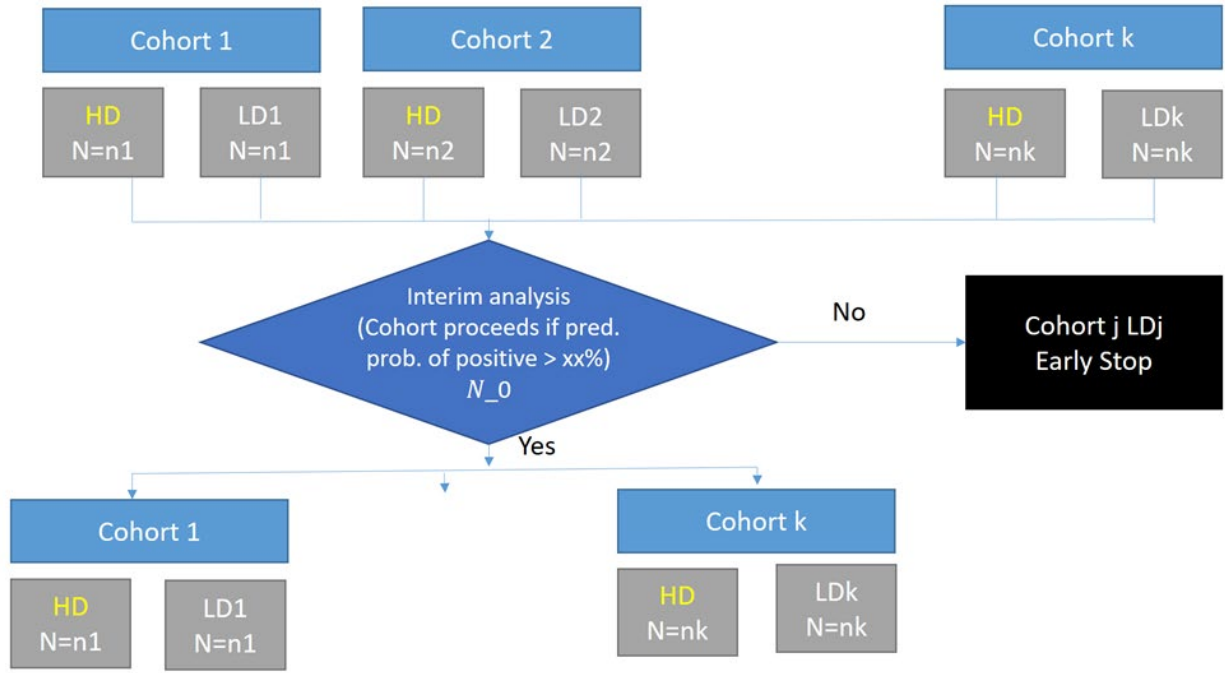

*Figure 5 The randomized fractional factorial design for dose optimization. HD, high dose; LD, low dose.*

**Figure 6** illustrates the outcomes of a hypothetical fractional factorial design with five cohorts and two RDEs within each cohort, where six different doses are compared. This fractional factorial design offers reductions of sample sizes when compared with the full factorial design. For example, if there are 20 patients enrolled for each RDE in each cohort, the commonly used full factorial design would require a total sample size of approximately 600 patients (5 × 6 × 20), but a fractional factorial design only requires around 200 patients (5 × 2 × 20). If this fractional factorial design is used for k cohorts, a total of k+1 different doses could be investigated in a randomized fashion for a higher chance of finding the optimal dose(s). Note that some of the low doses (LDs) could be the same and the full factorial design could be considered as a special case of the fractional factorial design with 100% as the fraction.

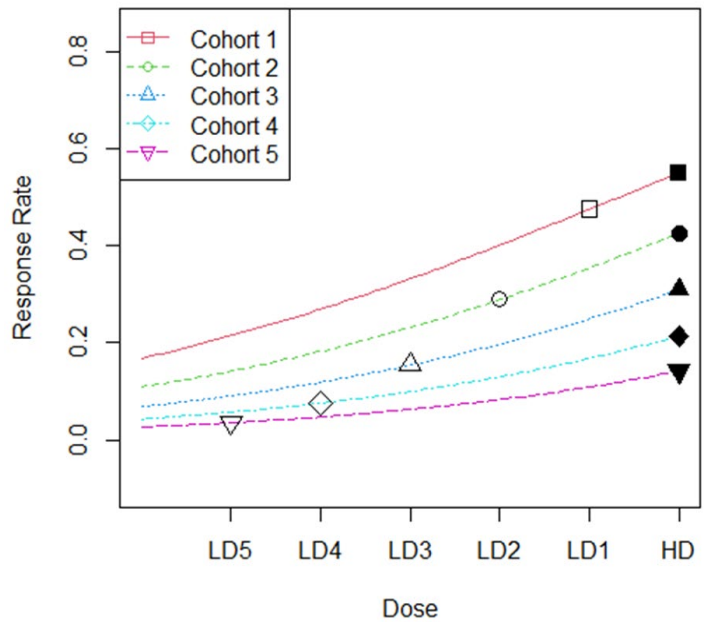

*Figure 6 Dose optimization design with five cohorts and different lower RDEs. HD, high dose; LD, low dose; RDEs, recommended doses for expansion.*

For optimal dose optimization, the recommended design is a fractional factorial design with all k+1 different doses investigated and compared with all cohorts together, instead of a full factorial design with only a subset of doses in each cohort where two doses were used in **Figure 5**. Note that, each cohort shares a common high dose (HD) to directly compare the efficacy and safety for tumor sensitivity. This fractional factorial design offers several advantages over a full factorial design: 1) the proposed design could efficiently investigate as many as k+1 dose levels for k different tumors and tumor sensitivity identified in a randomized fashion, which reduces the uncertainty of identifying the optimal dose and without extra cost; 2) exposure to suboptimal/toxic doses is well controlled and limited to a few patients in the staged design or continuous monitoring feature in an adaptive fashion; and 3) decision-making uses aggregated information from different cohorts. The fractional factorial design can be used for the expansion phase or standalone phase II studies. The sample size could be calculated with the desired efficacy in mind; as stated in the FDA dose optimization guidance, (FDA guidance, 2023) the chosen sample size is not necessarily powered for a certain hypothesis but rather, it is chosen for estimating the response rate with high precision. With an adequate sample size for each dose (RDE) studied, the general dose-response profile for the investigational treatment can be estimated efficiently.

With the efficacy and toxicity information from the randomized study, possible k+1 doses can be compared, where k is the number of cohorts. If the HD does not have any toxicity/tolerability concerns, then HD could be selected as the dose for the registration study and full assessment of efficacy and safety of the investigational treatment. If a toxicity concern arises with the HD, then one of the LDs could be chosen with a balanced benefit-risk profile, where a better toxicity profile is observed without much reduction in efficacy compared with the HD.

To demonstrate how the aggregated data of both efficacy and safety from the k cohorts could be used in decision-making, a statistical model can be built for inference. Let $L_i(\theta|D_i)$ be the likelihood from the i[th] cohort, i=1, ... , k, where θ are the parameters from the response model such as the logistic model and $D_i$ are the observed data from i[th] cohort. Thus, the likelihood function $L_0(\theta|D_1,\cdots,D_k)$ with all k cohorts from this randomized fractional factorial design can be obtained:

$$L_0(\theta|D_1,\cdots,D_k) \propto L_1(\theta|D_1)L_2(\theta|D_2)^{\alpha 2}\cdots L_k(\theta|D_k)^{\alpha k} \qquad (1)$$

Where $0 \leq \alpha j \leq 1$ (j=2, … , k) is a power parameter to discount the contribution from the $j^{th}$ cohort, $L_1(\theta|D_1)$ is the most sensitive tumor cohort with the highest response rate at the HD dose. When $\alpha j = 0$, it means the $j^{th}$ tumor cohort has no contribution to the most sensitive tumor evaluation. When $\alpha j = 1$, it indicates the $j^{th}$ tumor cohort has the full contribution, i.e., has the same response pattern as the most sensitive tumor. With the response data, these power parameters $\alpha j$ can be set as the ratio of the response rate from the $j^{th}$ tumor cohort to the response rate of the most sensitive tumor cohort 1. This setting is reasonable by calibrating and bringing all the tumor cohorts into the same level to make an inference. Note that the model (1) has the same format as the Bayesian power prior model (Ibrahim and Chen, 2000; Ibrahim, et al. 2015), borrowing information from different cohorts.

To assess the efficiency of this statistical inference with the aggregated data, the simulation study is used to show how well the selected optimal dose performs in comparison with the MTD/MAD, and the efficiency of the fractional factorial design against the full factorial design when the total number of patients is held constant. The goal of the simulation is first, to identify the most sensitive tumor cohort with the highest response rate at the HD, and second, to identify a dose that preserves the highest possible efficacy relative to that observed at the MTD/MAD.

Since one of the goals is to identify the tumor type most sensitive to the study treatment, it is desirable to estimate the dose-response in such tumor types. Using the statistical model, the lower confidence bound from the fitted value at the MTD/MAD is selected as the optimal dose, which does not have much efficacy lost compared with the MTD/MAD dose. Since typically the power parameters $\alpha j$ (j=2, ... , k) are unknown in model (1), the estimates using the observed response ratio of the j$^{th}$ tumor to the most sensitive tumor (i.e., the highest tumor response at the MTD/MAD) are used in the statistical inference.

*Table 1 Lower RDEs used for each cohort for different simulation settings.*

|  | Dose level (mg) | | | | |
| --- | --- | --- | --- | --- | --- |
|  | Cohort 1 | Cohort 2 | Cohort 3 | Cohort 4 | Cohort 5 |
| Scheme 1 | 250 | 300 | 350 | 400 | 450 |
| Scheme 2 | 450 | 400 | 350 | 300 | 250 |
| Scheme 3 | 450 | 300 | 250 | 350 | 400 |
| Scheme 4 | 250 | 400 | 450 | 350 | 300 |
| Scheme 5 | 250 | 300 | 350 | 400 | 450 |

RDEs, recommended doses for expansion

In the example simulation, a logistic dose-response is assumed, and that MTD/MAD is 500 mg and the HD for all five cohorts. Cohorts 1–5 range from the most sensitive tumor type to the least sensitive tumor type at the MTD, i.e., Cohort 1 has the highest tumor response at the MTD/MAD. A lower dose arrangement is assumed for each of Schemes 1–4 in the fractional factorial design as shown in **Table 1**. It is also assumed that there are 30 patients for each RDE dose in each cohort. Thus, the total number of patients is 300 (30 × 2 × 5). To demonstrate the efficiency of the fractional factorial design, a full factorial design with the same number of 300 patients is constructed as Scheme 5 with an equal number of patients in all six RDEs for each cohort. Thus, the number of patients in each RDE/cohort is 10 and the total number is the same as 300 (10 × 6 × 5).

*Table 2 The operating characteristics of the chosen optimal dose from the simulations.*

| Scheme | P(select) | CI | Estimated Dose | | | RR of Estimated Dose | | | |
|---|---|---|---|---|---|---|---|---|---|
| | | | Mean | Median | SD | Mean | Median | SD | % (RR<0.7) |
| 1 | 0.85 | 80% | 443.9 | 454.5 | 41.1 | 85.0 | 87.6 | 10.0 | 4.5 |
| | | 90% | 426.7 | 442.5 | 52.8 | 80.6 | 84.4 | 12.7 | 8.9 |
| | | 95% | 409.6 | 432 | 62.9 | 76.3 | 81.5 | 14.9 | 15.2 |
| 2 | | 80% | 474.3 | 475 | 6.0 | 93.0 | 93.2 | 1.6 | 0.0 |
| | | 90% | 467.4 | 468.5 | 9.7 | 91.1 | 91.4 | 2.5 | 0.1 |
| | | 95% | 460.6 | 463 | 17.5 | 89.3 | 89.9 | 4.3 | 0.5 |
| 3 | | 80% | 469.8 | 471 | 8.0 | 91.8 | 92.1 | 2.2 | 0.0 |
| | | 90% | 461.5 | 463.5 | 12.6 | 89.6 | 90.1 | 3.3 | 0.2 |
| | | 95% | 453.8 | 457 | 18.9 | 87.5 | 88.3 | 4.7 | 0.6 |
| 4 | | 80% | 448.8 | 458.5 | 40.0 | 86.3 | 88.7 | 9.7 | 3.6 |
| | | 90% | 431.9 | 447.5 | 52.8 | 82.0 | 85.7 | 12.7 | 7.6 |
| | | 95% | 413.5 | 437.5 | 65.6 | 77.4 | 83.0 | 15.6 | 13.9 |
| 5 | 0.73 | 80% | 450.5 | 455.5 | 25.4 | 86.6 | 87.9 | 6.3 | 1.2 |
| | | 90% | 438.4 | 446 | 34.3 | 83.4 | 85.3 | 8.3 | 2.7 |
| | | 95% | 427.1 | 438 | 43.3 | 80.5 | 83.1 | 10.3 | 5.1 |

CI, confidence interval; P(select), probability of selecting the most sensitive tumor cohort; RR, response rate; SD, standard deviation.

The response rate (RR) at each RDE dose for each tumor cohort was simulated according to the response curves for five different cohorts (**Figure 6**). This simulation was repeated 10,000 times for each scenario. **Table 2** summarizes the operating characteristics of the chosen optimal dose in terms of estimated optimal dose and the relative efficacy of the estimated dose to the MTD/MAD dose. P(select) is the probability of selecting the most sensitive tumor cohort (i.e., Cohort 1). Relative RR (%) is the relative efficacy of the estimated dose to the HD in Cohort 1 and is defined as $100 \times f_1(\widehat{D})/f_1(MTD)$, where $f_1(x)$ is the dose-response function, and $\widehat{D}$ is the chosen optimal dose. As shown in **Table 2**, when the total number of patients is held constant, the probability of selecting the most sensitive tumor cohort using the full factorial design in Scheme 5 is lower than using the fractional factorial designs (Schemes 1–4). This is due to the smaller number of patients allocated equally to each RDE with a larger variability in the full factorial design. Upon comparing the estimated dose from different design schemes, Scheme 2 has the best performance in terms of standard deviation for both estimate and relative efficacy of the estimated dose, followed by Scheme 3. As outlined in the results in **Table 2**, Scheme 2 may also lead to a better choice of LD for each cohort, when compared with other schemes. As there are different lower dose RDEs, it is preferable to put the highest LD RDE in the most hypothetically sensitive tumor cohort for better relative efficacy of the selected optimal dose.

3. **Summary**

We proposed a "three-step towards dose optimization" procedure (**Figure 7**) in response to the FDA's Project Optimus, which aligns with the recently published FDA Guidance for Industry for on dose optimization for drug development in oncology. (FDA guidance, 2023) The proposed three-step procedure is: 1) a dose-escalation part to identify MTD or MAD using an efficient hybrid design; 2) identification and exploration of multiple RDEs using all available data (such as pre-clinical data, the emerging clinical safety, PK, PD, and other biomarker information, for exposure-response model and efficacy/toxicity-response model for the dose-expansion phase); and 3) dose optimization using data from a randomized fractional factorial design with multiple different RDEs explored in the expansion phase or a phase II study to ensure that a feasible optimal dose is selected for registration trials.

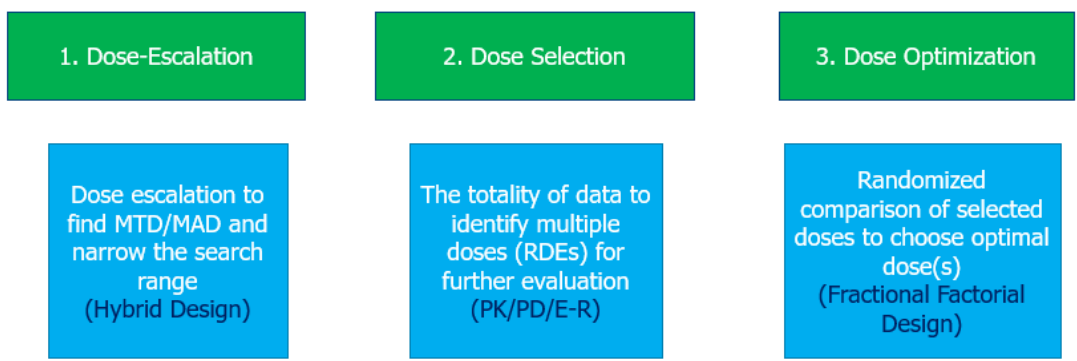

*Figure 7 A summary of the three-step approach leading to dose optimization.*

The efficient hybrid design in the dose-finding study offers control of overdose toxicities and can lead to an effective recommended MTD that is close to the true MTD. The randomized fractional factorial design for the dose optimization would enable efficient identification of drug-sensitive tumor types, while the use of adaptive design may help limit patients' exposure to suboptimal or toxic dose(s). As such, this three-step procedure is likely to select a recommended dose that retains most of the observed efficacy at the MTD/MAD and has a favorable safety profile. Of note, we recommend identifying at least 3 RDEs for comparisons in the third step, to reduce the uncertainty of optimal dose selection for drugs that may have non-linear dose-response profiles.

We also provided simulation data to support our recommendation for allocating a higher LD RDE to the patient cohort with the tumor type that is the hypothetically most sensitive to the investigative treatment for potential better efficacy outcomes.

The design for dose optimization (**Figure 5**) could have several stages, and it could be continuously monitored using the Bayesian predicted probability of success, for example, until certain futility criterial are met, the desired precision for response estimate is achieved, or allocated budget is maximized. With both safety and efficacy information from a randomized, multiple doses study of reasonable sample size investigated, an optimal dose could be chosen for discussion with a health agency regarding the design of a registration study. This three-step approach can be viewed as a seamless phase I/II study design for finding the optimal dose. If the investigational treatment is a combination of treatments, then a formal phase II study for the contribution of components could be carried out.

**Acknowledgments**

Editorial assistance was provided by Envision Pharma Group (Philadelphia, PA), and funded by Incyte Corporation, Wilmington, DE.

**References**

Araujo D et al. Oncology phase I trial design and conduct: time for a change – MDICT Guidelines 2022. *Ann Oncol* 2023;34(1):48–60. https://doi.org/10.1016/j.annonc.2022.09.158

Birri J, Puri-Lechner M, Hertzberg, A. Five Broad Implications for FDA's Project Optimus: How to Prepare for the Future of Oncology Trials. *Insights* 2022, https://www.hallorancg.com/2022/06/02/fda-project-optimus/

FDA Oncology Center of Excellence. Project Optimus - Reforming the dose optimization and dose selection paradigm in oncology. 2021. https://www.fda.gov/about-fda/oncology-center-excellence/project-optimus. 2021

Friends of Cancer Research. Optimizing dosing in oncology drug development. Friends of Cancer Research Annual Meeting 2021. https://friendsofcancerresearch.org/sites/default/files/2021-11/Optimizing_Dosing_in_Oncology_Drug_Development.pdf.

Ibrahim JG, Chen M-H. Power prior distributions for regression models. *Stat Sci* 2000;15(1):46–60.

Ibrahim JG, et al. The power prior: theory and applications. *Stat Med* 2015;34(28):3724–3749.

Liao JJZ, Zhou F, Zhou H, Petruzzelli L, Hou K, Asatiani E. A hybrid design for dose finding oncology clinical trials. *Int J Cancer* 2022;151(9):1602–1610. doi:10.1002/ijc.34203

Shah M, Rahman A, Theoret MR, et al. The drug-dosing conundrum in oncology: when less is more. *N Engl J Med* 2021;385:1445–1447.

Shah M, Rahman A, Theoret MR, Pazdur R. How to get the dose right. *The ASCO Post* https://ascopost.com/issues/may-10-2022/how-to-get-the-dose-right/
21

US Food and Drug Administration Guidance for Industry (2022). Expansion cohorts: use in first-in-human clinical trials to expedite development of oncology drugs and biologics guidance for industry. March 2022. https://www.fda.gov/media/115172/download.

US Food and Drug Administration (2022). Getting the dose right: optimizing dose selection strategies in oncology: an FDA-ASCO virtual workshop, May 3–5, 2022. https://www.fda.gov/news-events/fda-meetings-conferences-and-workshops/getting-dose-right-optimizing-dose-selection-strategies-oncology-fda-asco-virtual-workshop-05032022.

US Food and Drug Administration Guidance for Industry (2023). Optimizing the dosage of human prescription drugs and biological products for the treatment of oncologic diseases. January 2023. https://www.fda.gov/regulatory-information/search-fda-guidance-documents/optimizing-dosage-human-prescription-drugs-and-biological-products-treatment-oncologic-diseases

Zhou H, Zhou F, Liao JJZ. Package 'HybridDesign', 2022. https://cran.r-project.org/web/packages/HybridDesign/HybridDesign.pdf

Zirkelbach JF, et al. Improving dose-optimization processes used in oncology drug development to minimize toxicity and maximize benefit to patients. *J Clin Oncol* 2022;40(30):3489–3500.